\documentclass[a4paper,10pt,twoside]{cpc-hepnp}

\usepackage{multicol}
\usepackage{graphicx}
\usepackage{booktabs}
\usepackage{amssymb,bm,mathrsfs,bbm,amscd}
\usepackage[tbtags]{amsmath}
\usepackage{lastpage}

\begin{document}

\fancyhead[c]{\small Submitted to `Chinese Physics C'}
\fancyfoot[C]{\small 010201-\thepage}

\footnotetext[0]{Received 31 July 2014}

\title{Measurement of intrinsic radioactive backgrounds from the $^{137}$Cs and U/Th chains in CsI(Tl) crystals \thanks{Supported by the National Natural Science
Foundation of China (Contracts No. 11275107 and No. 11175099).}}

\author{
      S.K. Liu$^{1,2}$%
\quad Q. Yue$^{2;1)}$\email{yueq@mail.tsinghua.edu.cn}%
\quad S.T. Lin$^{1;2)}$\email{linst@phys.sinica.edu.tw}%
\quad Y.J. Li$^{2}$%
\quad C.J. Tang$^{1}$%
\quad H.T. Wong$^{3}$\\%
\quad H.Y. Xing$^{1}$%
\quad C.W. Yang$^{1}$%
\quad W. Zhao$^{2}$%
\quad J.J. Zhu$^{1}$%
}
\maketitle

\address{%
$^1$ College of Physical Science and Technology, Sichuan University, Chengdu, 610064\\
$^2$ Key Laboratory of Particle and Radiation Imaging (Ministry of Education) and\\ Department of Engineering Physics, Tsinghua University, Beijing, 100084\\
$^3$ Institute of Physics, Academia Sinica, Taipei, 11529\\
}

\begin{abstract}
The inorganic CsI(Tl) crystal scintillator is a candidate served as an anti-compton detector for China Dark matter EXperiment (CDEX). Studying the intrinsic radiopurity of CsI(Tl) crystal is an issue of major importance. The timing, energy and spatial correlations, as well as the capability of pulse shape discrimination provide powerful methods for the measurement of intrinsic radiopurities. The experimental design, detector performance and event-selection algorithms are described. A total of 359 $\times$ 3 kg-days data from three prototypes of CsI(Tl) crystals were taken at China Jinping underground laboratory where offers a good shielding environment. The contamination levels of internal isotopes from $^{137}$Cs, $^{232}$Th and $^{238}$U series, as well as the upper bounds of $^{235}$U series are reported. Identification of the whole $\alpha$ peaks from U/Th decay chains and derivation of those corresponding quenching factors are achieved.
\end{abstract}




\begin{multicols}{2}

\section{Introduction}

The CDEX pursues direct searches of light Weakly interactive massive particles (WIMPs) towards the goal of a ton-scale point-contact germanium detector array with sub-keV energy threshold \cite{2010_Kang,2012_Yue,2013_1kg,2013_Kang,2014_20g,2014_1kg}. A CsI(Tl) crystal array with the advantages of a relative high density, high light yield and low threshold is a good candidate for being the active (anti-compton) shielding of germanium detectors. Internal radioactive isotopes in CsI(Tl) crystal may contribute to the backgrounds of the germanium detectors. Quantitatively understanding of the intrinsic radioactive background is crucial. In addition to the active shielding of Ge detectors, the experiments which search for rare processes \cite{2000_Wong_AP} like dark matter \cite{2012_KIMS_PRL} or neutrino physics \cite{2003_Li_PRL,2010_Deniz_PRD} have to evaluate the contaminations of the internal isotopes to understand their contributions to the backgrounds.

This article focuses on the measurement of the intrinsic radiopurities of $^{137}$Cs, $^{238}$U, $^{232}$Th and $^{235}$U which are the major sources of the internal background in CsI(Tl). The data taking  spans over 79 days from Sep. 2012 to  Dec. 2012, providing the 359 kg-days of physics data for each of the three crystals.
The contaminations of U/Th series via the time correlated events from $\beta-\alpha$ and $\alpha-\alpha$ decay chains are derived.
In the following sections, the calibrations, pulse shape discrimination, the intrinsic radiopurites of $^{137}$Cs and U/Th series with time-correlation method are discussed. 

\section{Experimental setup}

The experiment was performed at the China Jinping Underground Laboratory (CJPL), which is located in one of several tunnels in Jinping mountain in Sichuan province.
With a rock overburden of more than 2400~m giving rise to a measured muon flux of 61.7 y$^{-1}\cdot$m$^{-2}$ \cite{2013_Kang_FP,2013_Wu_CPC}, this site provides an ideal location for low-background experiments, where the contribution of the cosmic ray background can be negligible. 

 A passive shielding system including, from outside to inside, 1 m of polyethylene, $\sim$15 cm of lead, $\sim$15 cm of OFHC (Oxygen-Free High-Conductivity) copper, was constructed. 
Three cuboid CsI(Tl) crystals with dimension of 60~mm $\times$ 60 mm $\times$ 280~mm are placed inside the shielding. 

\begin{center}
\includegraphics[width=1.0\linewidth]{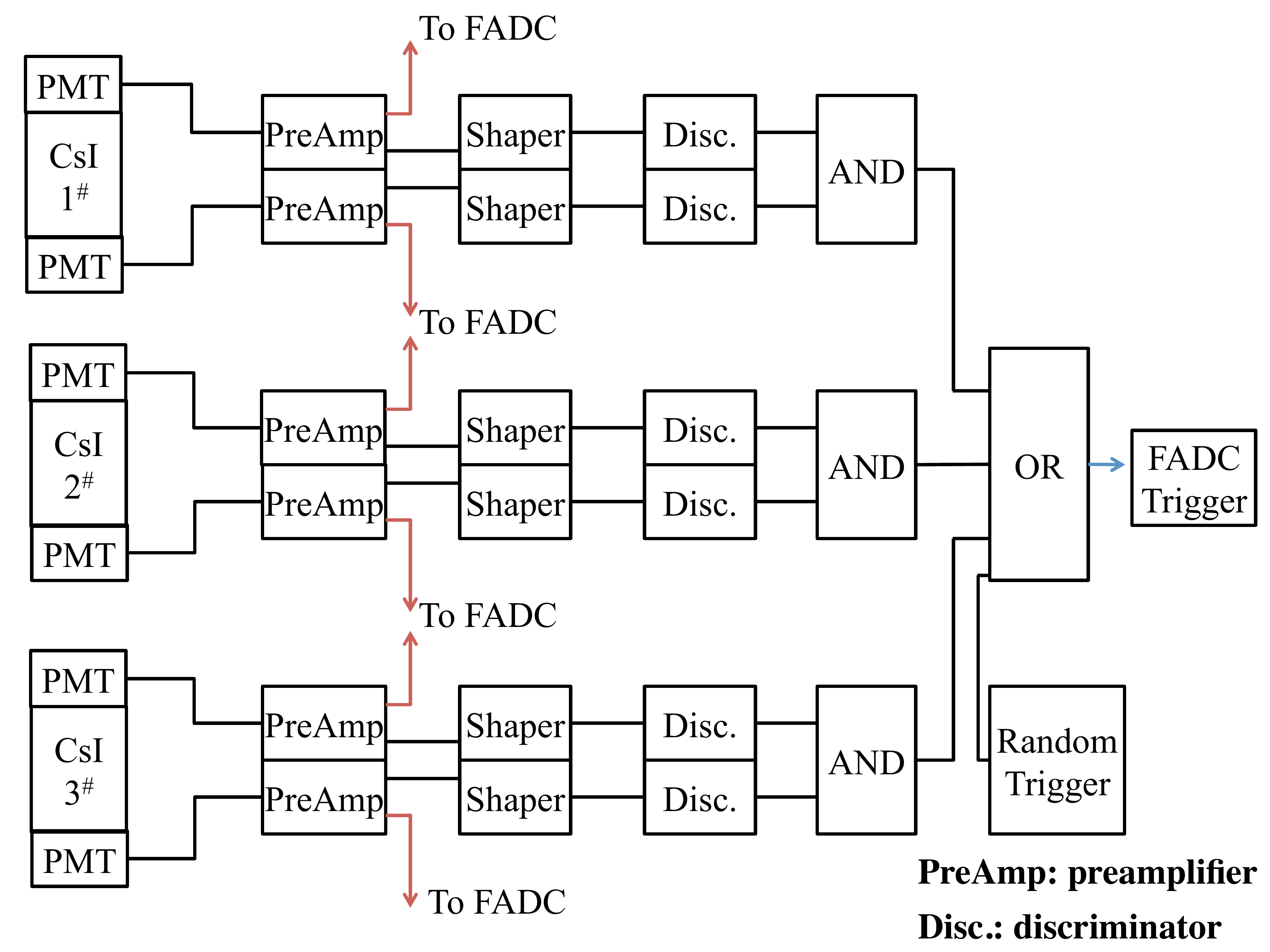}
\figcaption{\label{fig:daq}   Schematic diagram of the electronics and the DAQ system of three CsI(Tl) detectors. }
\end{center}

\begin{center}
\includegraphics[width=1.0\linewidth]{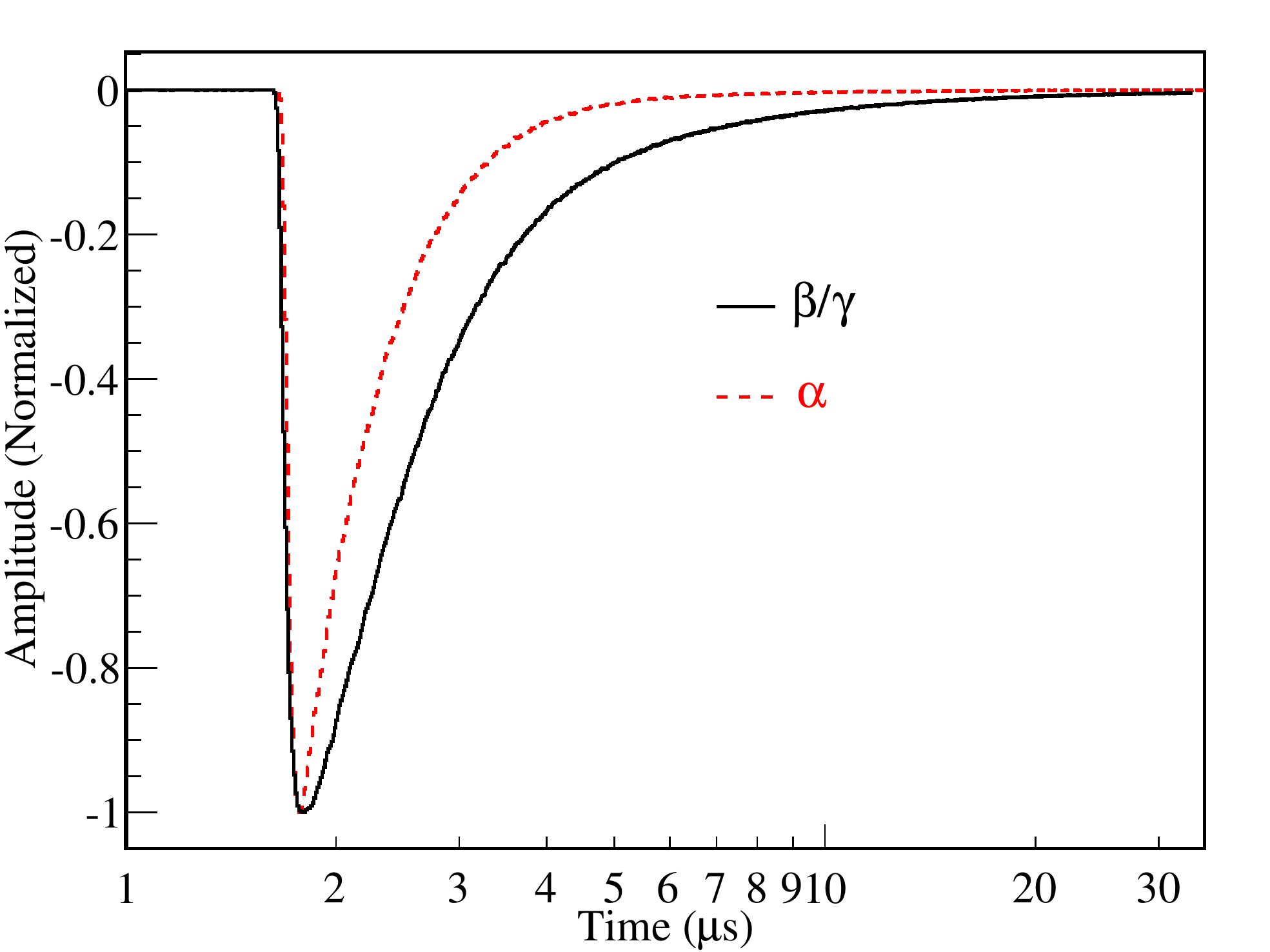}
\figcaption{\label{fig:pulse}   Averaged pulse shapes due to the $\alpha$ particles and $\beta/\gamma$ rays in the energy region of  2.6 MeVee (electron equivalent energy,``ee") to 2.7 MeVee are depicted, which are normalized by amplitudes. }
\end{center}

The schematic diagram of the electronics and data acquisition (DAQ) system is depicted in Fig.\ref{fig:daq}. The scintillation light is read out by photo-multipliers (PMTs), which have low radioactivity feature with free potassium borosilicate glass made by Beijing HAMAMATSU, at both ends of the crystal.
The PMT signal is read out by a preamplifier with two identical outputs. One is fed into a 100 MHz flash analog-to-digital convertor (FADC) where the signal is sampled and recorded. The other is fed into a shaper to generate triggers.
To reduce the accidental events from the noises of the PMTs,  the event is recored, only if both outputs from one crystal are triggered.
Events provided by a random trigger (RT) with a pulse generator at 0.05 Hz are recorded for calibration and DAQ dead time measurement. 
As long as the buffers of FADC are not full, the acquisition can continue without dead time in the following event. 
Thus, 99.9\% DAQ live time with a trigger rate of 6.2 Hz can be obtained.
The timing information of triggers is recorded by the FADC time tag with a resolution of 10 ns. The time correlations of the cascade events can be selected with significant background rejection.

\section{Calibrations and pulse shape discrimination}

The energy is defined by the sum of partial Q of the two PMTs. As shown in Fig.\ref{fig:pulse}, the partial Q is the integration of the pulse from 1 $\mu$s to 30 $\mu$s. The energy calibration is derived from the background events with explicit peaks from $^{137}$Cs, $^{40}$K, $^{214}$Bi,$^{208}$Tl and zero-energy defined by RT events, as depicted in the inset of Fig.\ref{fig:calibration}. The measured energy spectra of $\beta/\gamma+\alpha$ and  $\beta/\gamma$ with the explanations of the $\gamma$ peaks is shown in Fig.\ref{fig:calibration}.

\begin{center}
\includegraphics[width=1.0\linewidth]{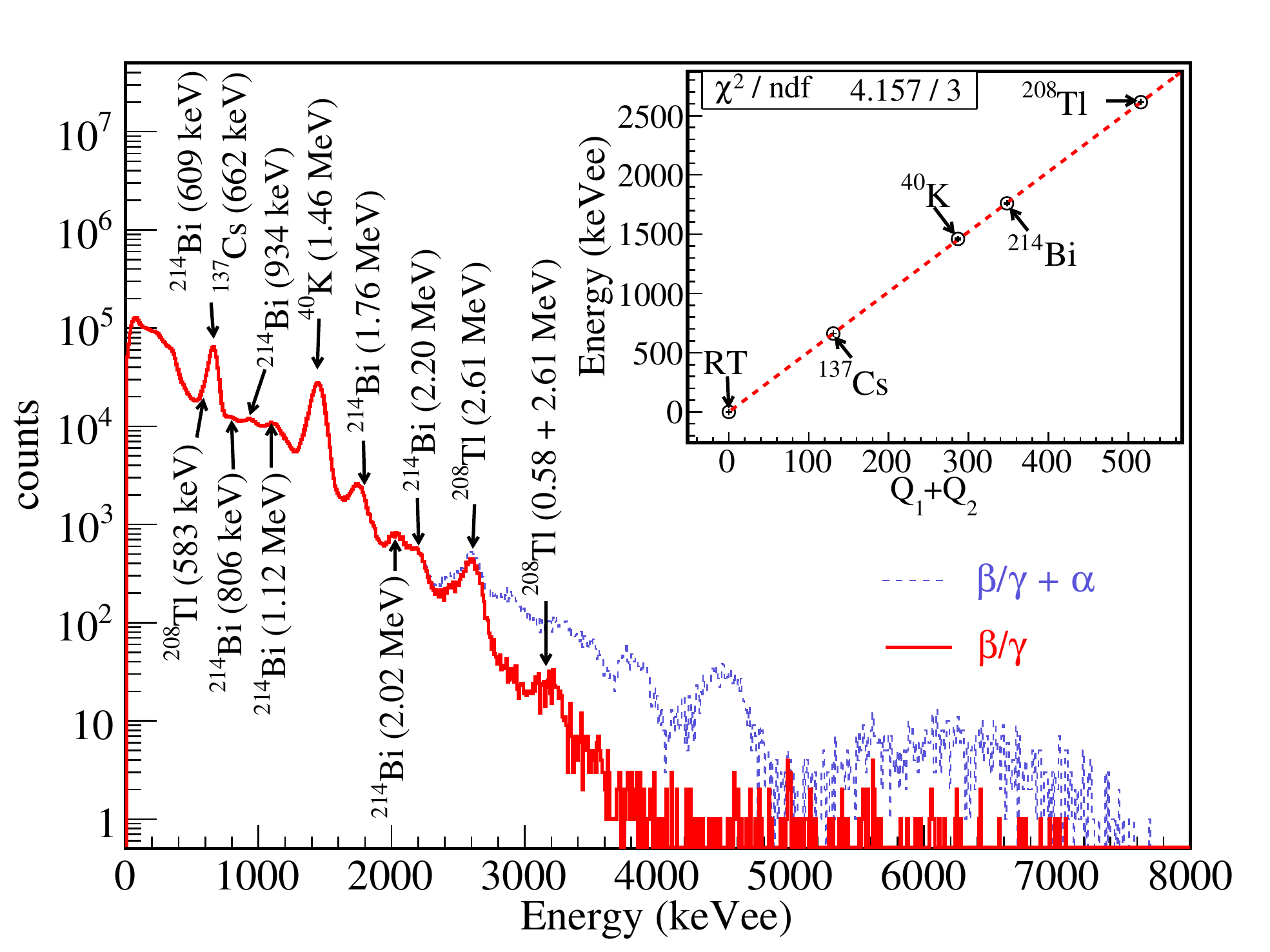}
\figcaption{\label{fig:calibration}  The measured energy spectra of $\beta/\gamma+\alpha$ and  $\beta/\gamma$, in which the $\gamma$ rays from U/Th decay chains and $^{137}$Cs, $^{40}$K are identified, as well as the linear calibration result shown in the inset. The error bars are smaller than the data point size.}
\end{center}

The CsI(Tl) crystal has the ability to discriminate the $\alpha$ events from the $\beta/\gamma$ events. The pulse shapes of heavily ionizing events from $\alpha$ particles have faster decay than those from $\gamma$ or electrons, as displayed in Fig.\ref{fig:pulse}. With the digitized pulse information recorded by FADC, the mean time method \cite{2004_PSD} is used for the pulse shape discrimination (PSD). The mean time is defined as
\begin{eqnarray}
\label{eq1}
\langle t\rangle = \frac{\sum_{i}A_{i}t_{i}}{\sum_{i}A_{i}}
\end{eqnarray}
where $A_{i}$ is the FADC amplitude at time-bin $t_{i}$.

\begin{center}
\includegraphics[width=1.0\linewidth]{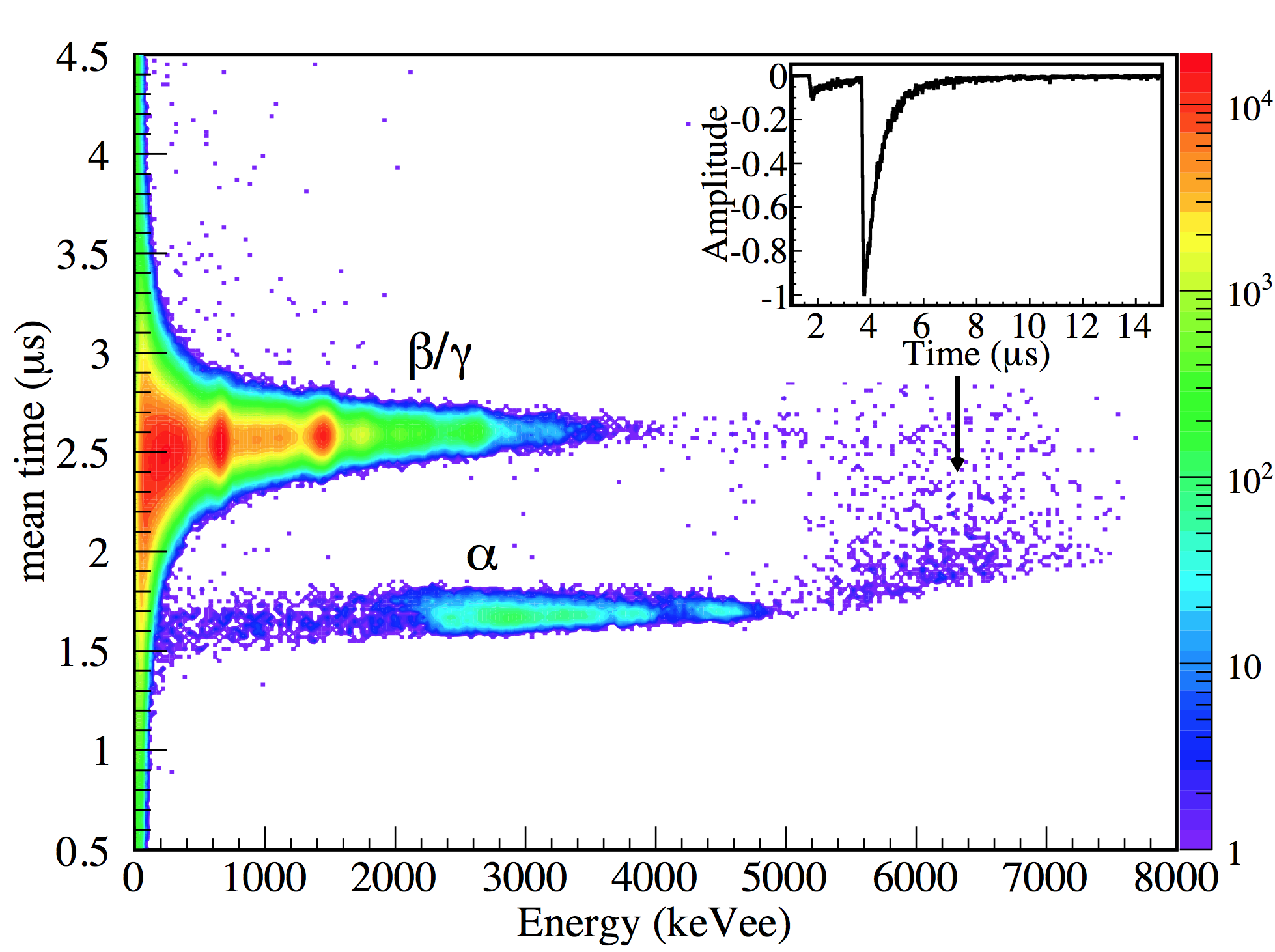}
\figcaption{\label{fig:meantime}  The correlation between mean time and measured energy for the crystal 1$^{\#}$. The background events are separated distinctly. The upper band is $\beta/\gamma$ events, the lower band is $\alpha$ events and the region above 5 MeVee is mainly due to $\beta/\gamma$ associated with the following $\alpha$ events. A pulse shape from those events is displayed in the inset.}
\end{center}

The correlation between the mean time and the measured energy is depicted in Fig.\ref{fig:meantime}. The $\beta/\gamma$ events and $\alpha$ events are separated into two distinct bands. The $\alpha$ pulses with faster falling tail are distributed in the lower band and the $\beta/\gamma$ events are distributed in the upper band. Due to the quenching factor effect \cite{1964_QF,1963_QF}, a majority of the alpha events from U/Th decay chains are ranged from 2 to 5 MeVee. In the energy region of less than $\sim$2 MeVee, the $\alpha$ events are attributed to ``surface alphas", which deposit only a partial of their energies in the CsI(Tl) crystal, because of the locations at the surface of a crystal. The half life of $^{212}$Po, which is the progeny of $^{232}$Th chain, is 0.3 $\mu$s that it associates with the $\beta$ decay from its parent $^{212}$Bi. 
\begin{eqnarray}
  ^{212}\text{Bi}&\to&^{212}\text{Po} + \overline{\upsilon}_{e} + e^{-} + \gamma^{'}s\nonumber\\
 &&(Q = 2.25 \text{MeV}, \tau_{1/2} = 60.6\text{ min})\nonumber\\
  ^{212}\text{Po}&\to&^{208}\text{Pb} + \alpha\nonumber\\
 &&(Q = 8.95 \text{ MeV}, \tau_{1/2} = \text{0.30 } \mu\text{s})\nonumber
\end{eqnarray}
The energy distribution of these events are more than 5 MeVee. The single $\alpha$ rate of three crystals above 300 eVee are 0.0014 kg$^{-1}$s$^{-1}$, 0.0007 kg$^{-1}$s$^{-1}$ and 0.0011 kg$^{-1}$s$^{-1}$, respectively, or one event per crystal per 158 s, 328 s and 203 s.

The selection of the high voltage (HV) of PMT has a prominent influence on the discrimination of $\alpha$ and $\beta/\gamma$. In order to quantify the discrimination, the figure-of-merit (FOM) parameter is defined as
\begin{eqnarray}
\label{eq2}
FOM = \frac{\left|M_{\gamma}-M_{\alpha}\right|}{\sigma_{\gamma}+\sigma_{\alpha}}
\end{eqnarray}

\begin{center}
\includegraphics[width=1.0\linewidth]{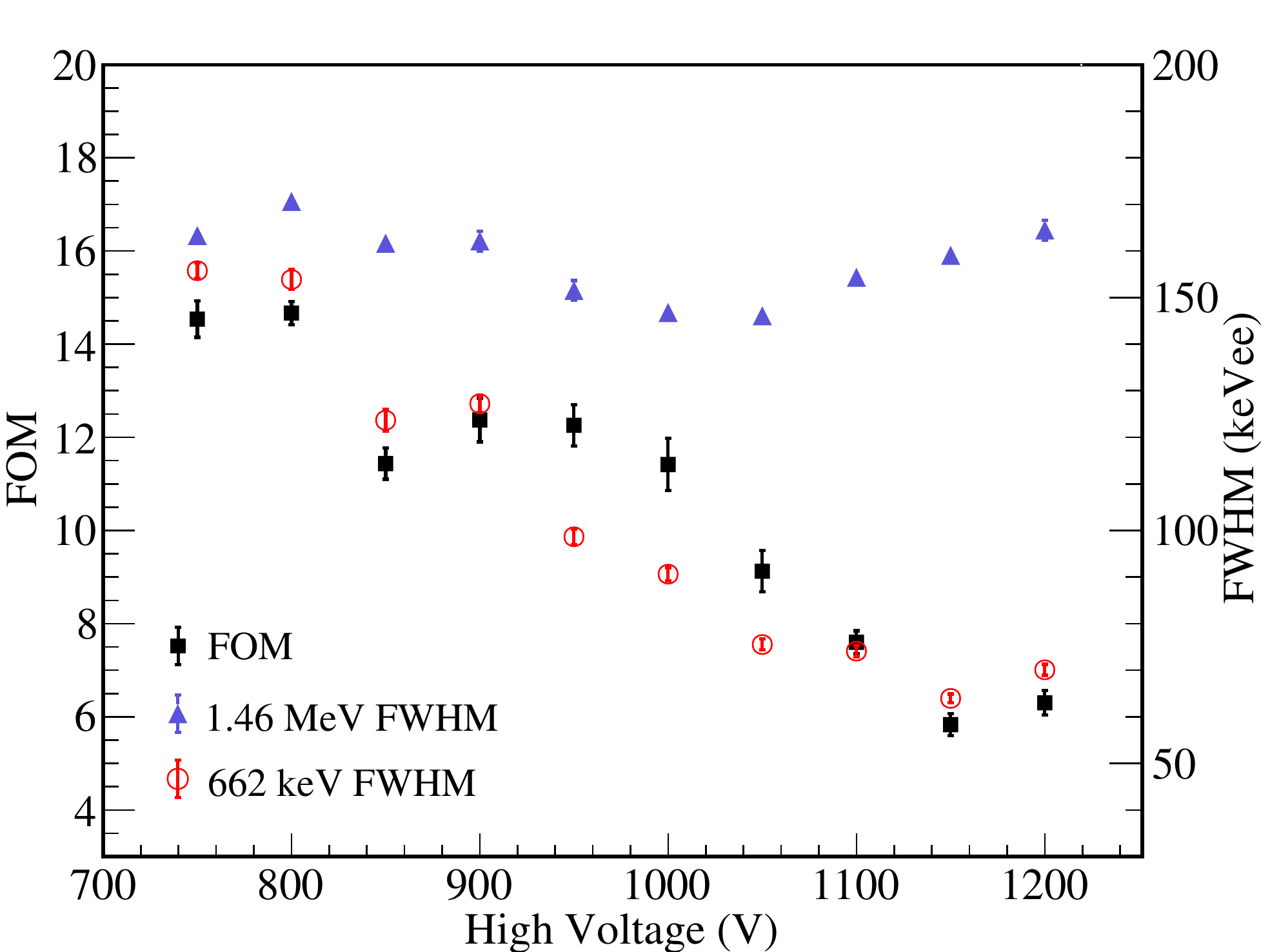}
\figcaption{\label{fig:FOM}  The correlation between the high voltage and the figure-of-merit representing the PSD ability of distinguishing $\alpha$ from $\beta/\gamma$ is depicted, as well as the full width at half maximum (FWHM) of 662 keVee and 1.46 MeVee from $^{137}$Cs and $^{40}$K. Some error bars are smaller than the data point size.}
\end{center}

where $M_{i}$ and $\sigma_{i}$  are the mean value and root-mean-square (RMS) of the $i (i=\gamma, \alpha)$ distribution of the mean time, respectively, at the energy of 2 MeVee to 5 MeVee. The relationship between the FOM parameter and HV of the PMT is displayed in Fig.\ref{fig:FOM}, as well as the energy resolutions at 662 keVee and 1.46 MeVee. 
The high HV deteriorates the PSD ability, which is against the energy resolution results. 
The measured energies of $\alpha$ events usually locate at the high energy region of above 2 MeVee, and the scintillation light excited by $\alpha$ has more fast component with the higher and more narrow pulse. For $\alpha$ events, the PMT may be saturated and even easier saturated than $\gamma$ events at the same measured energy with a high HV. 
The energies of $^{137}$Cs and $^{40}$K peaks are too low to reflect the PSD ability restricted by PMT saturation.
However there are no intensive $\gamma$ peaks above 2 MeVee, so the FOM parameter is chosen as a criterion of the HV selection. A HV at $\sim$800 voltage demostrated the best PSD ability is selected.

The longitudinal (z) position can be derived by the variation of the dimensionless ratio $R=(Q_{1}-Q_{2})/(Q_{1}+Q_{2})$, where $Q_{1}$ and $Q_{2}$ are the partial Q of the PMT signals from two ends. The z position can be calibrated by the $^{40}$K peak, which is mainly from the materials of the PMTs in the vicinity of the CsI(Tl) crystal.

\section{Intrinsic radiopurity of $^{137}$Cs}

$^{137}$Cs is one of the dominant internal background sources in CsI(Tl) crystals and it is produced artificially as fission waste from power reactors and atomic weapon tests. $^{137}$Cs can be easily introduced during the processing from cesium ore, such as the water used for Cs extraction might be the main source for $^{137}$Cs contamination \cite{2005_Cs137}. 

\vspace{0.5em}
$^{137}$Cs is a beta decay via

\vspace{0.5em}
\leftline{$^{137}$Cs $\to$ $^{137}$Ba$^{*}$ + $\overline{\upsilon}_{e}$ + $e^{-}$ ($t_{1/2}$ = 30.1 y)}

\vspace{0.5em}
\leftline{$^{137}$Ba$^{*}$ $\to$ $^{137}$Ba + $\gamma$ ($t_{1/2}$ = 2.55 min, $E_{\gamma}$ = 662 keVee)}

\vspace{0.5em}
The emitted $\beta$ and subsequent $\gamma$ ray are not prompt. Therefore the 662 keVee $\gamma$ can be measured in isolation and evaluated the internal radiopurity of $^{137}$Cs with detector efficiency. 
An efficiency of 39.9\% for full energy deposition of the 662 keVee $\gamma$ rays in a single crystal is derived from the simulation. Owing to the minority contribution of 609 keVee $\gamma$ rays from $^{214}$Bi to the 662 keVee peak, the conservative radiopurity results are derived. The activities of three crystals are $86.6\pm0.1$ mBq kg$^{-1}$, $72.9\pm0.1$~mBq~kg$^{-1}$ and $74.5 \pm 0.1$ mBq kg$^{-1}$, respectively, or equivalently contamination levels of $2.85\pm0.03 \times10^{-17}$ g/g, $2.40\pm0.03 \times10^{-17}$  g/g and $2.45\pm0.03 \times10^{-17}$ g/g. These are similar to those from the previous results \cite{2003_Cs137_contamination,2006_Cs137_contamination}.

\section{Intrinsic radiopurities of U/Th series}

The $^{235}$U, $^{238}$U and $^{232}$Th are naturally occurring radioactive elements, and exist in all materials at a certain level. The progenies of U/Th series can emit $\gamma$ rays, which easily propagate and transmit energy to the Ge detector. The ICP-MASS analysis \cite{2000_ICPMASS} is a way to determine the mass concentrations of U/Th series, but it is difficult to lower its sensitivity below the order of ppt. However the 
timing-correlation method aiming to count the $\alpha$ events by itself is a relatively easy way to improve the sensitivity \cite{2003_Cs137_contamination,2006_Cs137_contamination}. This method can only give the contamination levels of the progenies emitting $\alpha$ particles. If secular equilibrium is assumed, the radiopurities of $^{235}$U, $^{238}$U and $^{232}$Th can be derived by the measured activities of those progenies.

\subsection{Decay sequences selection}

The U/Th decay sequences involve many sequential $\alpha$ and $\beta$ decays until they finally reach the stable isotopes. Among these sequences, the cascade decays with $\alpha$ emitting and their half lives longer than the data taking window but shorter than the rate of $\alpha$ event are considered. 

As to $^{238}$U series, the $\beta-\alpha$ cascade (C$_{1}$) is selected:
\begin{eqnarray}
  ^{214}\text{Bi}&\to&^{214}\text{Po} + \overline{\upsilon}_{e} + e^{-} + \gamma^{'}s\nonumber\\
 &&(Q = 3.28 \text{MeV}, \tau_{1/2} = 19.9 \text{ min})\nonumber\\
  ^{214}\text{Po}&\to&^{210}\text{Pb} + \alpha\nonumber\\
 &&(Q = 7.83 \text{ MeV}, \tau_{1/2} = \text{164 } \mu\text{s})\nonumber
\end{eqnarray}
The contamination of $^{226}$Ra with $\tau_{1/2}$=1600 y can be derived by the event rate of this cascade.

As to $^{232}$Th series, the $\alpha-\alpha-\alpha$ cascade (C$_{2}$) is selected:
\begin{eqnarray}
  ^{224}\text{Ra}&\to&^{220}\text{Rn} + \alpha\nonumber\\
 &&(Q = 5.79 \text{ MeV}, \tau_{1/2} = 3.66 \text{ d})\nonumber\\
  ^{220}\text{Rn}&\to&^{216}\text{Po} + \alpha\nonumber\\
 &&(Q = 6.41 \text{ MeV}, \tau_{1/2} = \text{55.6 }\text{s})\nonumber\\
   ^{216}\text{Po}&\to&^{212}\text{Pb} + \alpha\nonumber\\
 &&(Q = 6.91 \text{ MeV}, \tau_{1/2} = \text{0.145 }\text{s})\nonumber
\end{eqnarray}
The event rate of this cascade determines the radiopurity of $^{228}$Th with  $\tau_{1/2}$=1.91 y.

As to $^{235}$U series, the $\alpha-\alpha-\alpha$ cascade (C$_{3}$) is selected:
\begin{eqnarray}
  ^{223}\text{Ra}&\to&^{219}\text{Rn} + \alpha\nonumber\\
 &&(Q = 5.97 \text{ MeV}, \tau_{1/2} = 11.4 \text{ d})\nonumber\\
  ^{219}\text{Rn}&\to&^{215}\text{Po} + \alpha\nonumber\\
 &&(Q = 6.94 \text{ MeV}, \tau_{1/2} = \text{3.96 }\text{s})\nonumber\\
   ^{215}\text{Po}&\to&^{211}\text{Pb} + \alpha\nonumber\\
 &&(Q = 7.52 \text{ MeV}, \tau_{1/2} = \text{1.78 }\text{ms})\nonumber
\end{eqnarray}

The event rate of this cascade gives the radiopurity of $^{227}$Ac with  $\tau_{1/2}$=21.8 y.

\subsection{Event selections}

We carried out the measurement of timing-correlation method to evaluate the U/Th chain concentrations in CsI(Tl) crystal. A number of data analysis selections were adopted in order to select the candidate cascade events, as well as their corresponding signal efficiencies evaluation. The selections are discussed as follows, meanwhile the results are summarized in Table.~\ref{table:selections}.

\begin{enumerate}
\item
The PSD selection: The events were firstly identified as $\alpha$ or $\beta/\gamma$ by the mean time method shown in Fig.~\ref{fig:meantime}.

\item
The z position selection ($Z$): The density plots of the z position versus measured energy from the $\alpha$ events, as well as the $Z$ cut, are depicted in Fig.~\ref{fig:z_cut} (a). The distribution of the z position at both ends is distortion. In addition, it is easier for both ends to expose to the ambient air, because of the untighted wrapping at both ends of the crystal,  where $^{222}$Rn was not purged by nitrogen gas. $^{222}$Rn is a noble gas that permeates the air and its progeny $^{218}$Po is a reactive metal which readily adheres to almost any surface \cite{2000_Rn222_PLB}. It turns out the extra background of our measurement. Furthermore, the PMTs, which were in the vicinity of both ends, may also give extra $\alpha$ events there. In order to get rid of the influence on the end effects and obtain better $\alpha$ resolutions, the events from the middle section of the crystal are retained. The $\alpha$ spectra of crystal 1$^{\#}$ with and without Z selection are displayed in Fig.~\ref{fig:z_cut} (b).

\begin{center}
\includegraphics[width=1.0\linewidth]{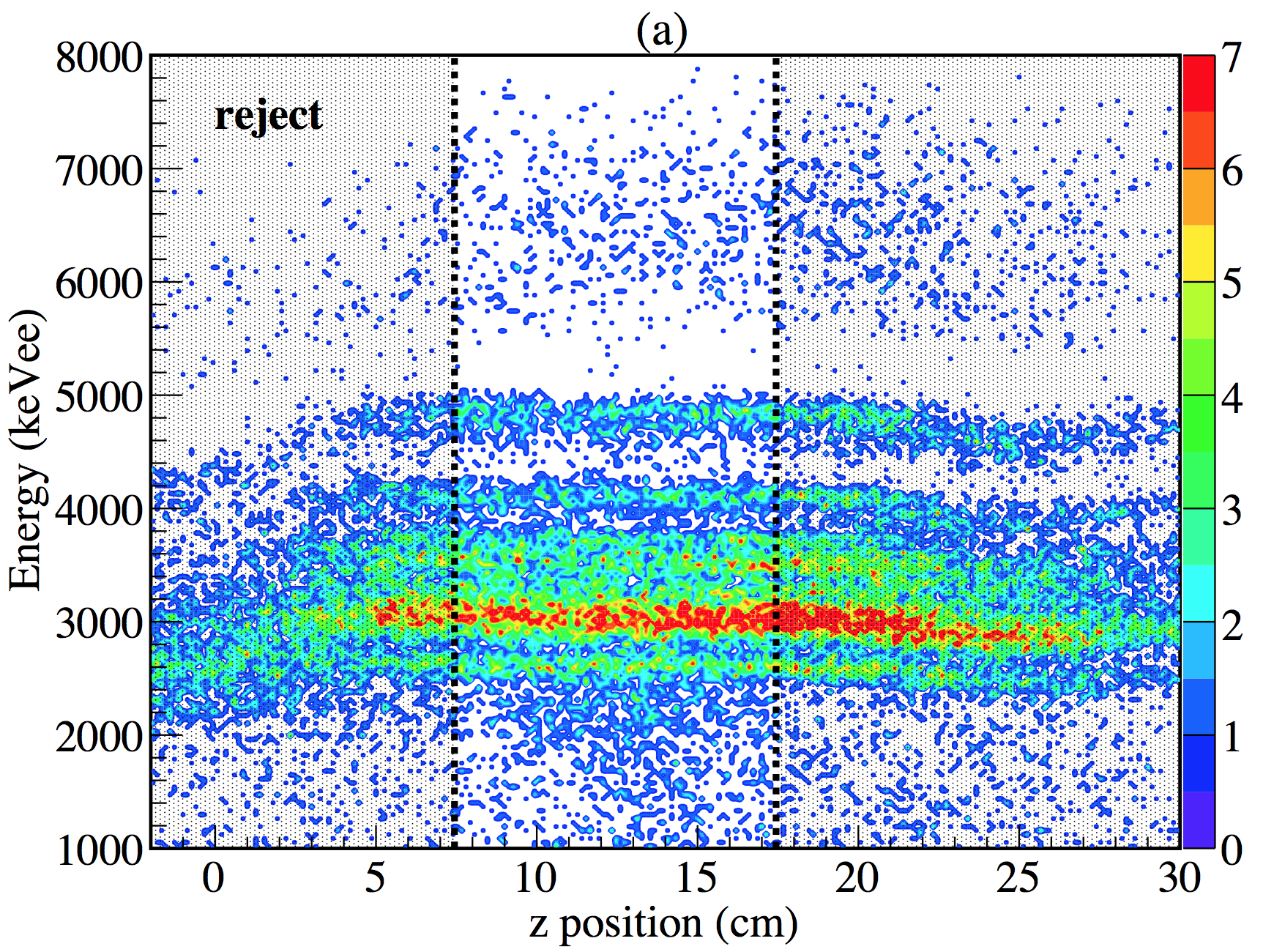}
\includegraphics[width=1.0\linewidth]{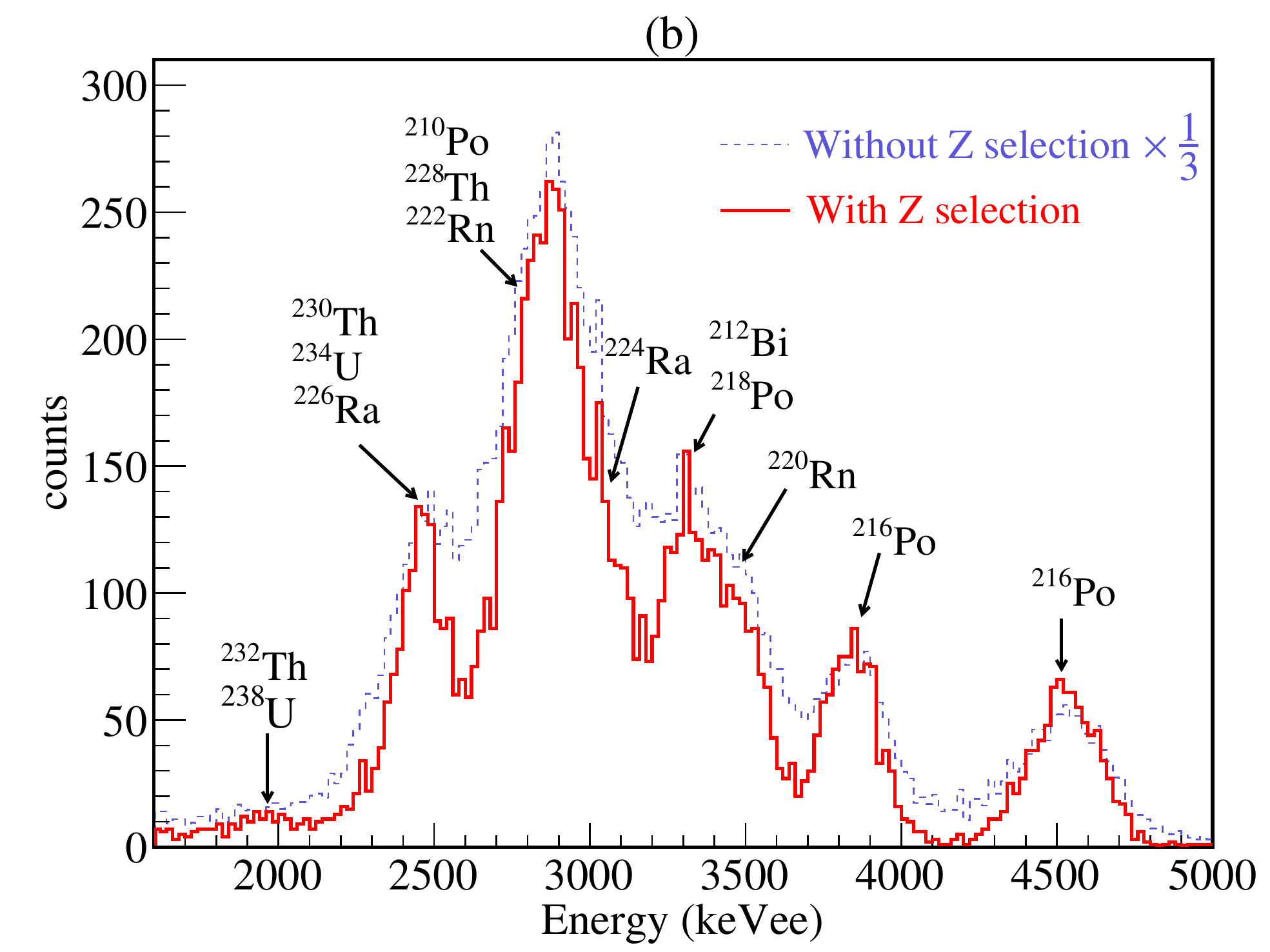}
\figcaption{\label{fig:z_cut}  (a) Scatter plots of the z distribution versus energy from $\alpha$ events of crystal 3$^{\#}$. The events in the black shade are rejected. (b) The $\alpha$ spectrum of crystal 1$^{\#}$ with Z selection is depicted as well as the explanations of the compositions of the $\alpha$ peaks. In addition, the $\alpha$ spectrum after scaled without Z selection is also shown to compare. }
\end{center}

\item
The half life selection ($\Delta t$): For each $\alpha$ event, the time correlations between this $\alpha$ event and the previous (later) $\alpha$ ($\beta/\gamma$) event were recorded. The time difference between the selected  decay sequence were required to be less than 5 times the nominal half lives and be at least more than one time window width.

\item
The energy selection ($E$): 
For the $\alpha$ event in the related cascade, the energy selection was within $\pm3\sigma$ region of the Gaussian mean value of the measured energy. While that of the $\beta$ events ($^{214}\text{Bi}
\to^{214}\text{Po}$) was above 100 keVee due to the the energy threshold of approximately 100 keV.
\item
The $\Delta z$ selection ($\Delta z$): The $\Delta z$ distribution of correlated $\beta-\alpha$ and $\alpha-\alpha$ cascades are depicted in Fig.~\ref{fig:Delta_z}. The $\alpha-\alpha$ correlated cascade provide a more accurate description ($\sigma_{\Delta z} = 1.86$ cm) in the spatial resolution, since both of the $\alpha$ pairs are deposited energy at the almost same site. While the $\beta-\alpha$ correlated cascade has a worse accurate spatial resolution ($\sigma_{\Delta z} = 3.30$ cm), because the accompanying $\gamma$ rays may not deposit energy at the original site of the $\beta$ decay. Nevertheless, the events within the $\pm3\sigma$ region of the Gaussian distribution of the z difference between the correlated cascade decay were selected.

\end{enumerate}

The final numbers of $\alpha-\alpha$ and $\beta-\alpha$ decays are obtained by counting the numbers of events after applying all selections above. Besides, the accidental events, which are  due to uncorrelated event-pairs which survive the various selections and fall into time windows corresponding to five half-lives, are statistically subtracted. 

\vspace{0mm}
\begin{center}
\tabcaption{ \label{table:selections}  Summary of the event selection procedures of crystal 1$^{\#}$ at two decay sequences from $^{232}$Th and $^{238}$U respectively. Listed are the individual and cumulative background survival fraction [$\lambda$(\%) and $\Pi \lambda$(\%), respectively] and the candidate signal efficiency [$\epsilon$(\%)].}
\footnotesize
\begin{tabular*}{75mm}{c@{\extracolsep{\fill}}ccc}
 \toprule Decay sequence&$^{214}\text{Bi}\to^{214}\text{Po}$&$^{220}\text{Rn}\to^{216}\text{Po}$\\
 \hline
  Raw counts	&  \multicolumn{2}{c}{7212712}  \\ 
  \hline
  \multicolumn{3}{c}{PSD selection$^{\dagger}$}\\
    	\multicolumn{1}{c}{$\lambda [ \Pi \lambda ]$(\%)}&\multicolumn{2}{c}{0.60[0.60]} \\
	\multicolumn{1}{c}{$\epsilon$(\%)}	& \multicolumn{2}{c}{100}\\
  \hline
  \multicolumn{3}{c}{Z selection} \\
    	\multicolumn{1}{c}{$\lambda [ \Pi \lambda ]$(\%)} & \multicolumn{2}{c}{25.9[0.16]} \\
	\multicolumn{1}{c}{$\epsilon$(\%)}	& \multicolumn{2}{c}{42.9} \\
  \hline
  \multicolumn{3}{c}{$\Delta t$ selection} \\
    	\multicolumn{1}{c}{$\lambda  [ \Pi \lambda ]$(\%)}	&8.2[0.01]& 8.6[0.01]\\	
	\multicolumn{1}{c}{$\epsilon$(\%)}	& 77.8 & 96.9\\
  \hline
  \multicolumn{3}{c}{E selection} \\
    	\multicolumn{1}{c}{$\lambda  [ \Pi \lambda ]$(\%)}	&79.2[0.01]& 90.4[0.01]\\	
	\multicolumn{1}{c}{$\epsilon$(\%)}	& 96.2& 99.4\\
  \hline
  \multicolumn{3}{c}{$\Delta z$ selection} \\
    	\multicolumn{1}{c}{$\lambda  [ \Pi \lambda ]$(\%)}	&96.3[0.01]& 99.8[0.01]\\	
	\multicolumn{1}{c}{$\epsilon$(\%)}	& 99.7 & 99.7\\
  \hline
  Coincidence counts & 0.06 & 0.14 \\
  Final counts & 705 & 872\\
  Measured activity& &\\
   (mBq kg$^{-1}$) &  \raisebox{1.0ex}[0pt]{0.071$\pm$0.003} & \raisebox{1.0ex}[0pt]{0.068$\pm$0.002} \\
\bottomrule
\multicolumn{3}{l}{$^\dagger$ Select $\alpha$ events above 300 eVee.} \\
\end{tabular*}
\vspace{0mm}
\end{center}

\vspace{0mm}
\begin{center}
\includegraphics[width=1.0\linewidth]{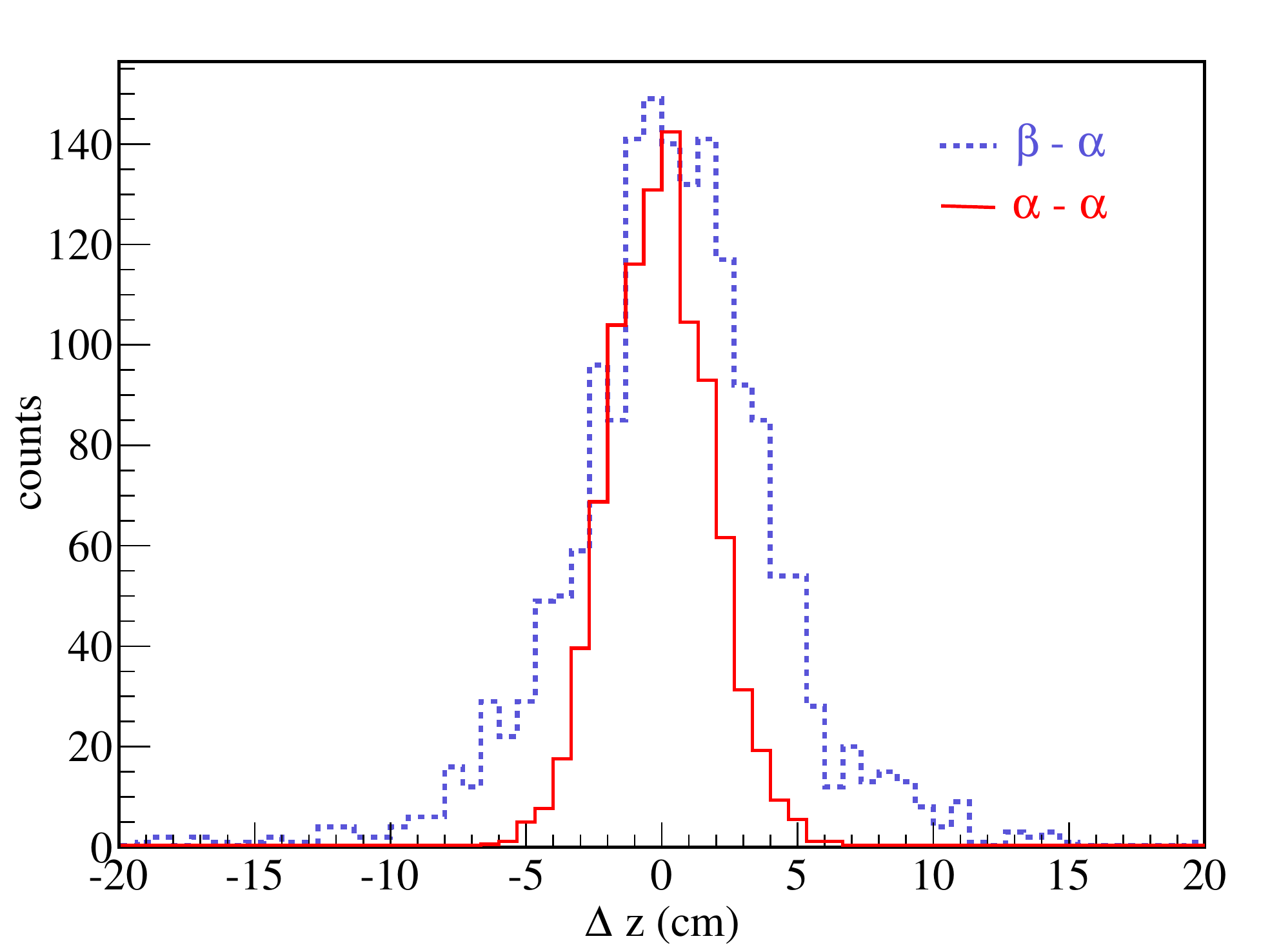}
\figcaption{\label{fig:Delta_z}  The $\Delta z$ distributions of $\beta-\alpha$ cascade decay ($^{214}\text{Bi}\to^{214}\text{Po}$) and $\alpha-\alpha$ cascade decay ($^{220}\text{Rn}\to^{216}\text{Po}$).}
\end{center}
\vspace{0mm}

\subsection{Efficiency corrections}

The efficiency corrections are considered by four factors. The first $\epsilon_{Z}$ is the correction of the $Z$ selection, which is the ratio of the longitudinal length of the selected z to the total longitudinal length of the crystal. 

The second $\epsilon_{\Delta t}$ considering the events lost due to the half life selection is determined by 
\begin{eqnarray}
\label{eq3}
\epsilon_{\Delta t} = \frac{\int_{50 \mu s}^{5\tau_{1/2}}exp(-ln2/\tau_{1/2}t)dt}{\int_{0}^{\infty}exp(-ln2/\tau_{1/2}t)dt}
\end{eqnarray}

\begin{center}
\includegraphics[width=1.0\linewidth]{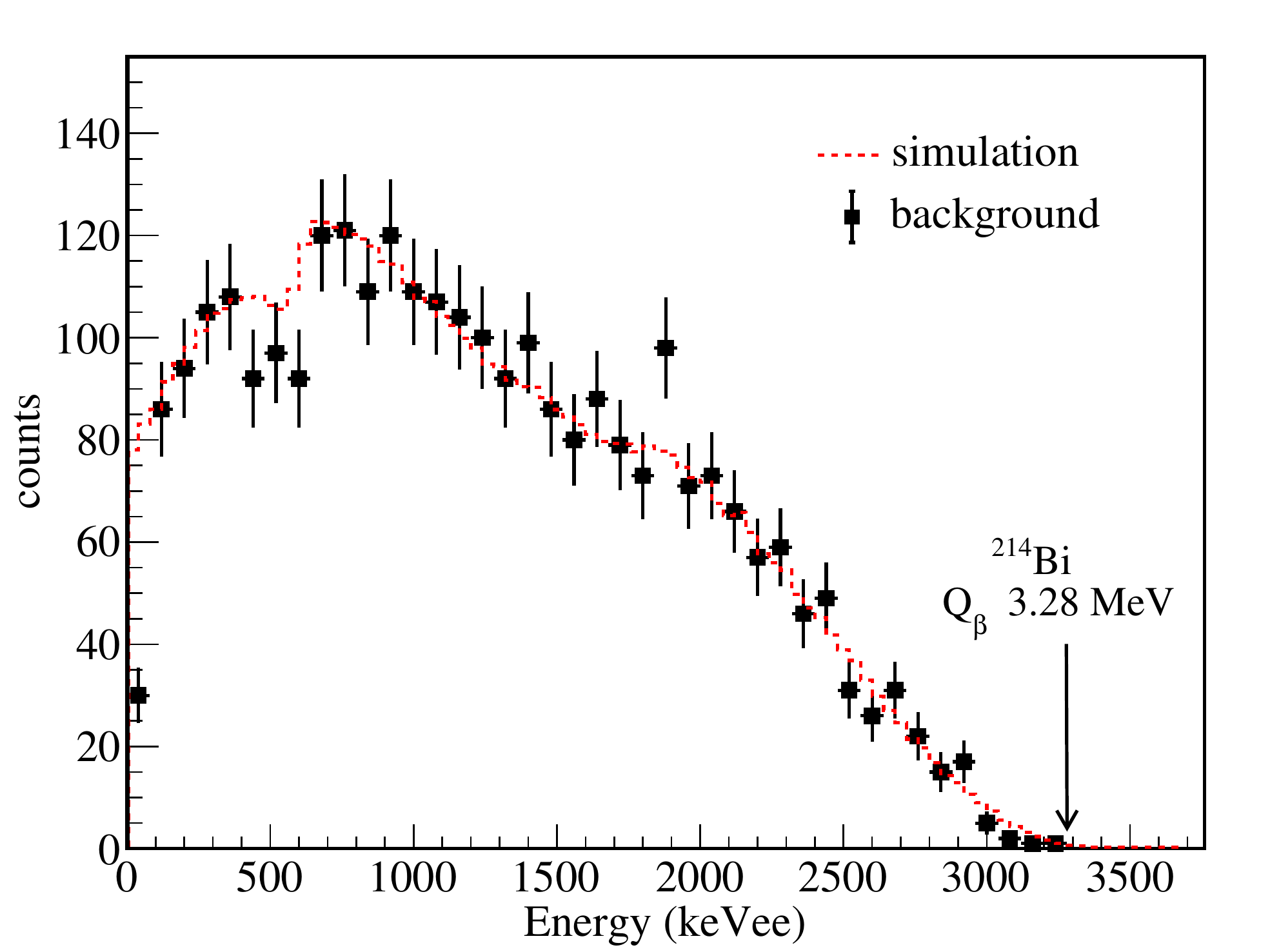}
\figcaption{\label{fig:Bi214_sim_exp}  The energy spectrum of $^{214}$Bi $\beta/\gamma$, as well as the Monte Carlo simulation result. They are in good agreement.}
\end{center}

The third $\epsilon_{E}$ has two conditions. For alpha events, $\epsilon_{E_{\alpha}}$ is the survival fraction within the $\pm3\sigma$ region of Gaussian distribution to be 99.7\%. For $\beta/\gamma$ events ($^{214}\text{Bi}
\to^{214}\text{Po}$), the survival fraction of events above 100 keVee ($\epsilon_{E_{\beta}}$) was derived from the simulation to be 96.5~\%.  The measured $\beta/\gamma$ spectrum, as well as the simulation result, is displayed in Fig.\ref{fig:Bi214_sim_exp}. The measured spectrum was well reproduced by the simulation. 

The fourth $\epsilon_{\Delta z}$, just like the $\epsilon_{E_{\alpha}}$, equals to 99.7\%, which is the percentage of events within the three standard deviations of Gaussian distribution.

The selection efficiencies, events suppression factors and accidental counts for each cascade decay at crystal 1$^{\#}$ are summarized in Table.~\ref{table:selections} to illustrate the effects of each process.

\subsection{Measurement results}

In order to establish the level of alpha events and cross check the correctness of the previous event selections, the $\Delta t$ distributions after applying all selections were fitted with the function below.
\begin{eqnarray}
\label{eq4}
f(\Delta t) \approx C(m + \lambda)\cdot e^{-(m + \lambda)\Delta t}
\end{eqnarray}
where $\lambda$ = ln2/$\tau_{1/2}$ and $m$ is the alpha event rate in one crystal.

For $^{214}$Po and $^{216}$Po, due to the short $\Delta t$ range, the fitting function can be simplified to 
\begin{eqnarray}
\label{eq4}
f(\Delta t) \approx C\lambda\cdot e^{-\lambda \Delta t}
\end{eqnarray}

The $\Delta t$ distributions for the correlated events from the crystal 1$^{\#}$ are shown in Fig.~\ref{fig:delta_t}. The measured half-lives ($\tau_{1/2}$) from all three crystals are summarized in Table.~\ref{table:half_life}. All the measurements are in excellent agreement with the standard values.

\begin{center}
\tabcaption{ \label{table:half_life}  Measured half lives of the decay sequences for all crystals.}
\footnotesize
\begin{tabular*}{83mm}{@{\extracolsep{\fill}}cccc}
 \toprule Decay &  & Measured & Nominal\\
 sequence & \raisebox{1.5ex}[0pt]{Detector} &  half life ($\tau_{1/2}$) & half life ($\tau_{1/2}$)\\
 \hline
  & crystal 1$^{\#}$ & 161$\pm$4 $\mu$s\\
  & crystal 2$^{\#}$ & 170$\pm$8 $\mu$s\\
  \raisebox{3ex}[0pt]{$^{214}\text{Bi}\to^{214}\text{Po}$}  & crystal 3$^{\#}$ & 167$\pm$5 $\mu$s & \raisebox{3ex}[0pt]{164 $\mu$s}  \\
 \hline
  & crystal 1$^{\#}$ &  0.143$\pm$0.005 s\\
  & crystal 2$^{\#}$ &  0.150$\pm$0.007 s\\
  \raisebox{3ex}[0pt]{$^{220}\text{Rn}\to^{216}\text{Po}$} & crystal 3$^{\#}$ &  0.144$\pm$0.004 s & \raisebox{3ex}[0pt]{0.145 s}\\
 \hline
  & crystal 1$^{\#}$ & 55.2$\pm$1.8 s\\
  & crystal 2$^{\#}$ & 54.5$\pm9.8$ s\\
  \raisebox{3ex}[0pt]{$^{224}\text{Ra}\to^{220}\text{Rn}$} & crystal 3$^{\#}$ & 54.6$\pm$2.8 s & \raisebox{3ex}[0pt]{55.6 s}\\
\bottomrule
\end{tabular*}
\end{center}

The measured contamination levels of three crystals are summarized in Table.~\ref{table:contaminations} in several ways, respectively. The radiopurity levels of the long-lived parents are measured directly from the corresponding cascade decay. Assuming secular equilibrium, the contamination levels of $^{238}$U and $^{232}$Th can also be derived.

\end{multicols}

\begin{center}
\tabcaption{ \label{table:contaminations}  Measured activities of all crystals as well as the derived contaminations.}
\footnotesize
\begin{tabular*}{160mm}{@{\extracolsep{\fill}}ccccc}
 \toprule Decay &  & Measured activity & Contamination of & Contamination of \\
 sequence & \raisebox{2ex}[0pt]{Detector} & (mBq kg$^{-1}$) & long-lived parents (g/g) & series (g/g) $^\dagger$\\
 \hline
  & crystal 1$^{\#}$ & 0.071$\pm$0.003 & $^{226}$Ra: 19.4$\pm$0.8$\times$10$^{-19}$ & $^{238}$U: 5.71$\pm$0.24$\times$10$^{-12}$\\
  & crystal 2$^{\#}$ & 0.022$\pm$0.002 & $^{226}$Ra: 6.29$\pm$0.55$\times$10$^{-19}$ & $^{238}$U: 1.85$\pm$0.16$\times$10$^{-12}$\\
  \raisebox{3ex}[0pt]{$^{214}\text{Bi}\to^{214}\text{Po}$} & crystal 3$^{\#}$ & 0.087$\pm$0.003 & $^{226}$Ra: 23.8$\pm$0.8$\times$10$^{-19}$ & $^{238}$U: 6.99$\pm$0.24$\times$10$^{-12}$\\
 \hline
  & crystal 1$^{\#}$ & 0.068$\pm$0.002& $^{228}$Th: 22.4$\pm$0.7$\times$10$^{-22}$ & $^{232}$Th: 16.7$\pm$0.5$\times$10$^{-12}$\\\
  & crystal 2$^{\#}$ & 0.017$\pm$0.002& $^{228}$Th: 5.60$\pm$0.66$\times$10$^{-22}$ & $^{232}$Th: 4.17$\pm$0.49$\times$10$^{-12}$\\\
  \raisebox{3ex}[0pt]{$^{220}\text{Rn}\to^{216}\text{Po}$}& crystal 3$^{\#}$ & 0.069$\pm$0.003& $^{228}$Th: 22.7$\pm$0.8$\times$10$^{-22}$ & $^{232}$Th: 16.9$\pm$0.6$\times$10$^{-12}$\\
\bottomrule
\multicolumn{5}{l}{$^\dagger$ Assuming secular equilibrium.} \\
\end{tabular*}
\end{center}

\begin{center}
\tabcaption{ \label{table:U235_contaminations}  Measured limit sensitivities of $^{235}$U for all crystals as well as the derived contaminations.}
\footnotesize
\begin{tabular*}{160mm}{@{\extracolsep{\fill}}ccccc}
 \toprule Decay &  & Limit sensitivity & Contamination of & Contamination of \\
 sequence & \raisebox{2ex}[0pt]{Detector} & (mBq kg$^{-1}$) & long-lived parents (g/g) & series (g/g) $^\dagger$\\
 \hline
  & crystal 1$^{\#}$ & $1.1\times10^{-3}$& $^{227}$Ac: $<4.1\times10^{-22}$& $^{235}$U: $<1.4\times10^{-14}$\\
  & crystal 2$^{\#}$ & $5.5\times10^{-4}$& $^{227}$Ac: $<2.1\times10^{-22}$& $^{235}$U: $<6.9\times10^{-15}$\\
  \raisebox{3ex}[0pt]{$^{219}\text{Rn}\to^{215}\text{Po}$}& crystal 3$^{\#}$ & $2.1\times10^{-4}$& $^{227}$Ac: $<7.8\times10^{-23}$& $^{235}$U: $<2.6\times10^{-15}$\\
\bottomrule
\multicolumn{5}{l}{$^\dagger$ Assuming secular equilibrium.} \\
\end{tabular*}
\end{center}

\begin{multicols}{2}

\begin{center}
\includegraphics[width=1.0\linewidth]{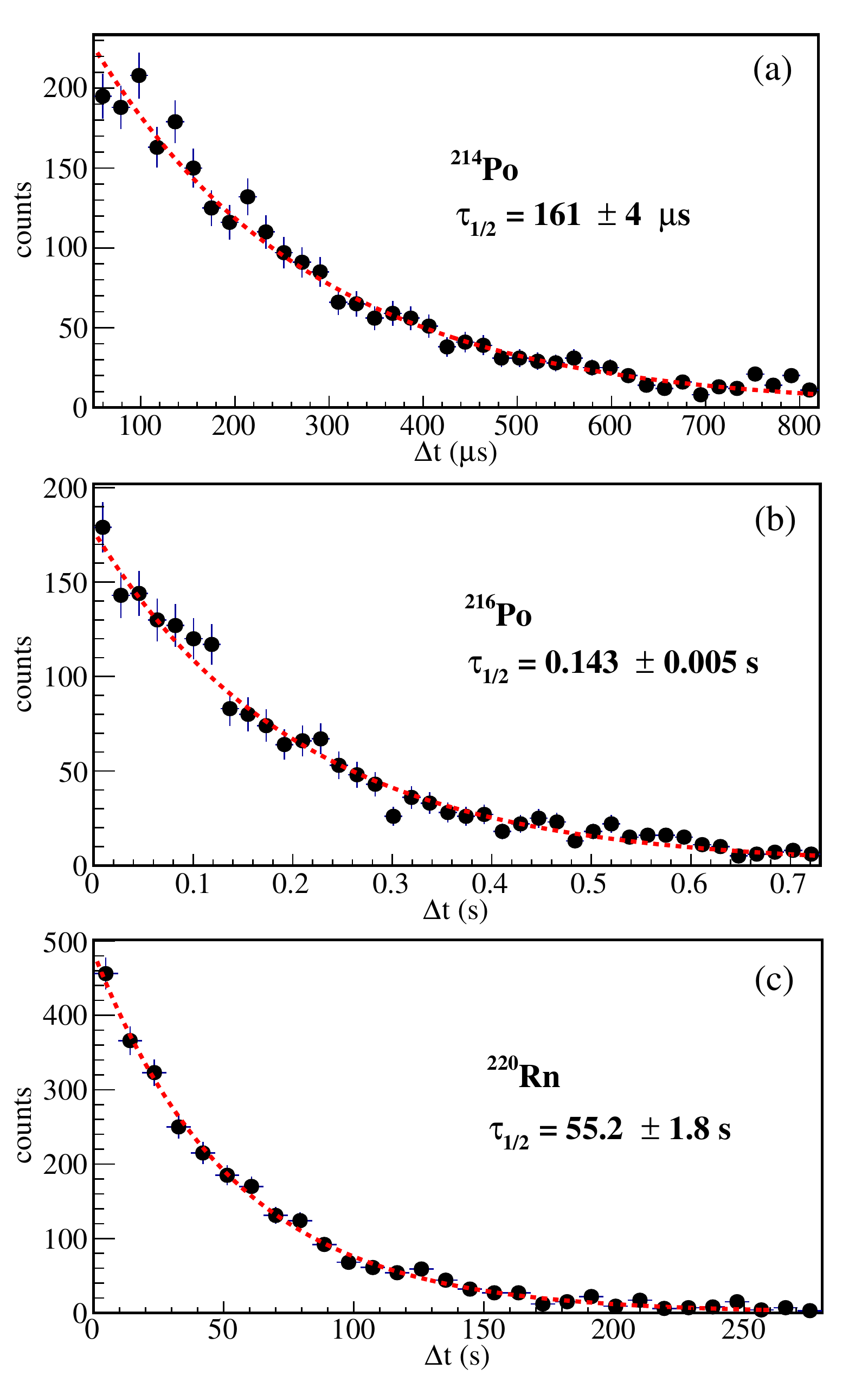}
\figcaption{\label{fig:delta_t}  The $\Delta t$ distributions representing the half lives of $^{214}$Po (a), $^{216}$Po (b) and $^{220}$Rn (c) are depicted. The Eq. (5) is applied to fit the half lives of  $^{214}$Po and $^{216}$Po, while the Eq. (4) is used to derive that of $^{220}$Rn, where $m$ is the background alpha rate subtracted the $^{216}$Po $\alpha$ rate. }
\end{center}

For $^{235}$U decay chain shown in Fig.~\ref{fig:U235_delta_t}, no time-correlated events are observed in delay time ($\Delta t$) distribution of $\alpha-\alpha$ events for $^{219}\text{Rn}\to^{215}\text{Po}$ from crystal 1$^{\#}$. Because of the absence of the correlated events, the upper limit converted from the accidental rate is derived to be $1.1\times10^{-3}$~mBq kg$^{-1}$, corresponding to $4.1\times10^{-22}$~g/g of $^{227}$Ac radiopurity level and $1.4\times10^{-14}$~g/g of $^{235}$U contamination level assuming secular equilibrium. The measured intrinsic contamination levels of the other two crystals are also listed in Table.~\ref{table:U235_contaminations}. 

Amount of light produced in scintillating material by alphas is lower than that produced by electrons at the same energy. Thus, the $\alpha$ events are observed at lower energies than their real values with the calibration done by electron or $\gamma$ sources. The quenching factor (QF) of $\alpha$ particles is defined as
\begin{eqnarray}
\label{eq4}
QF = \frac{E_{ee}}{E_{Q}}
\end{eqnarray}
where $E_{ee}$ is the measured ``electron-equivalent" energy of the $\alpha$ particles and $E_{Q}$ is the Q-value of the $\alpha$ decays. The value of $E_{Q}$ is exactly known and the $E_{ee}$ can be derived by the identified $\alpha$ events using the time-correlated cascade decay.

The QF of crystal 2$^{\#}$ is shown in Fig.~\ref{fig:QF}, as well as the result from reference \cite{2006_Cs137_contamination}. Those of the other two crystals are summarized in Table.~\ref{table:QF}. The differences of those results are probably because of the different amount of Tl dopant and the time window in which scintillation signal is collected \cite{2010_QF_AP}.

\begin{center}
\tabcaption{ \label{table:QF}  Measured quenching factors for all three crystals.}
\footnotesize
\begin{tabular*}{80mm}{@{\extracolsep{\fill}}ccccc}
 \toprule E$_{Q}$ & \multicolumn{3}{c}{Quenching Factor}   \\ \cline{2-4}
 (MeV) & crystal 1$^{\#}$ &crystal 2$^{\#}$&crystal 3$^{\#}$\\
 \hline
 5.79  & 0.5268$\pm$0.0017 & 0.5060$\pm$0.0017&0.5596$\pm$0.0017\\
 6.41 & 0.5490$\pm$0.0016 & 0.5526$\pm$0.0016 & 0.5788$\pm$0.0016\\
 6.91 &  0.5542$\pm$0.0014& 0.5355$\pm$0.0014 & 0.5933$\pm$0.0013\\
 7.83 & 0.5773$\pm$0.0013 & 0.5505$\pm$0.0013 & 0.6194$\pm$0.0013 \\
\bottomrule
\end{tabular*}
\end{center}

\begin{center}
\includegraphics[width=1.0\linewidth]{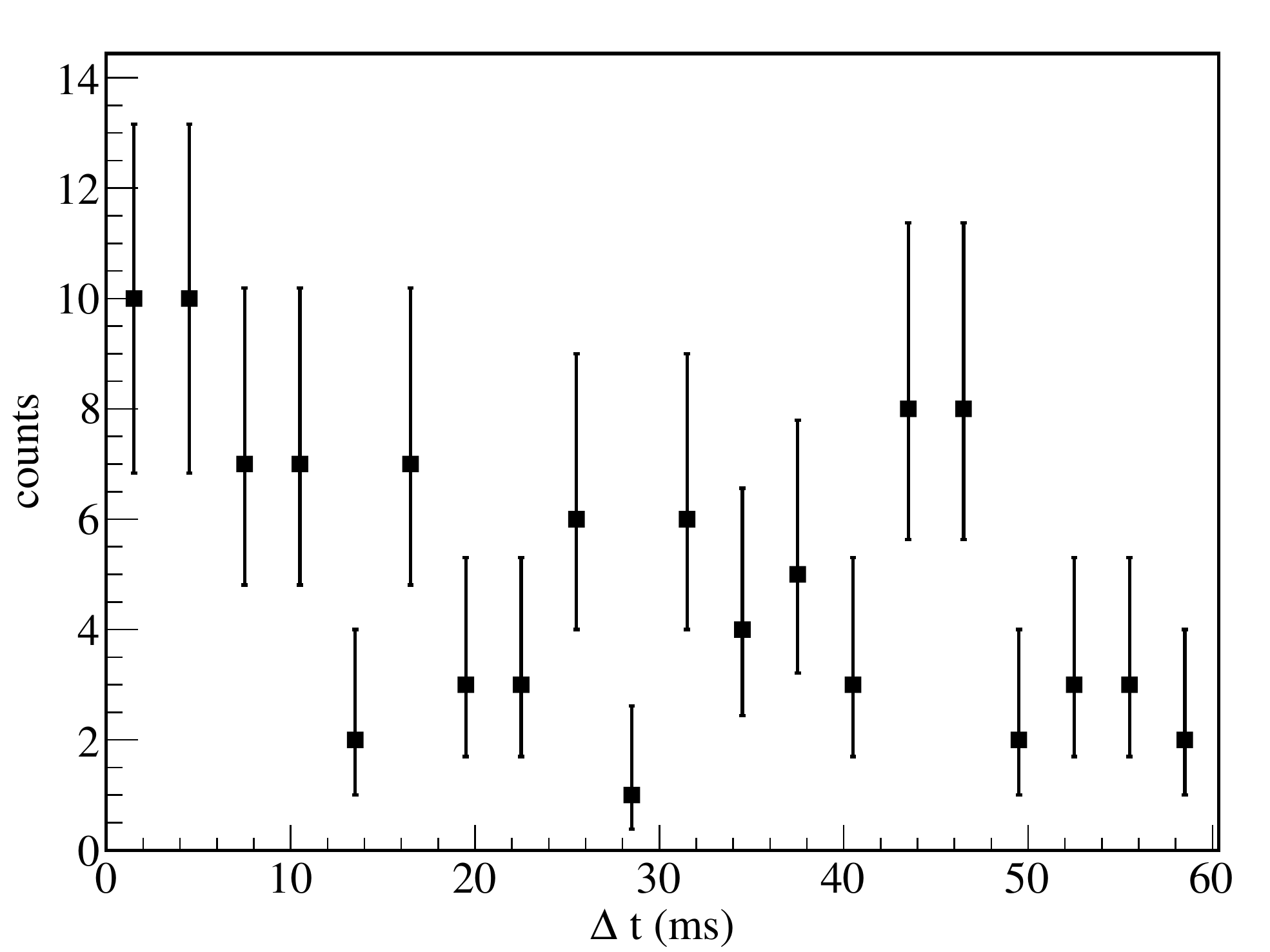}
\figcaption{\label{fig:U235_delta_t}  The $\Delta t$ distribution of  $\alpha-\alpha$ events for $^{219}\text{Rn}\to^{215}\text{Po}$ from crystal 1$^{\#}$. No correlated events are observed in such that only the upper limit converted from the accidental rate can be derived. }
\end{center}

\section{Summary and conclusion}

In this work, the experiment of three CsI(Tl) crystals was performed at CJPL with good shielding system. 
The calibrations of CsI(Tl) detectors were derived from the background $\gamma$ rays, as well as the quenching factors of $\alpha$ particles from the background $\alpha$ events.
The PSD ability has been studied using mean time method, in addition to optimizing the high voltage of PMTs via the PSD ability.

\begin{center}
\includegraphics[width=1.0\linewidth]{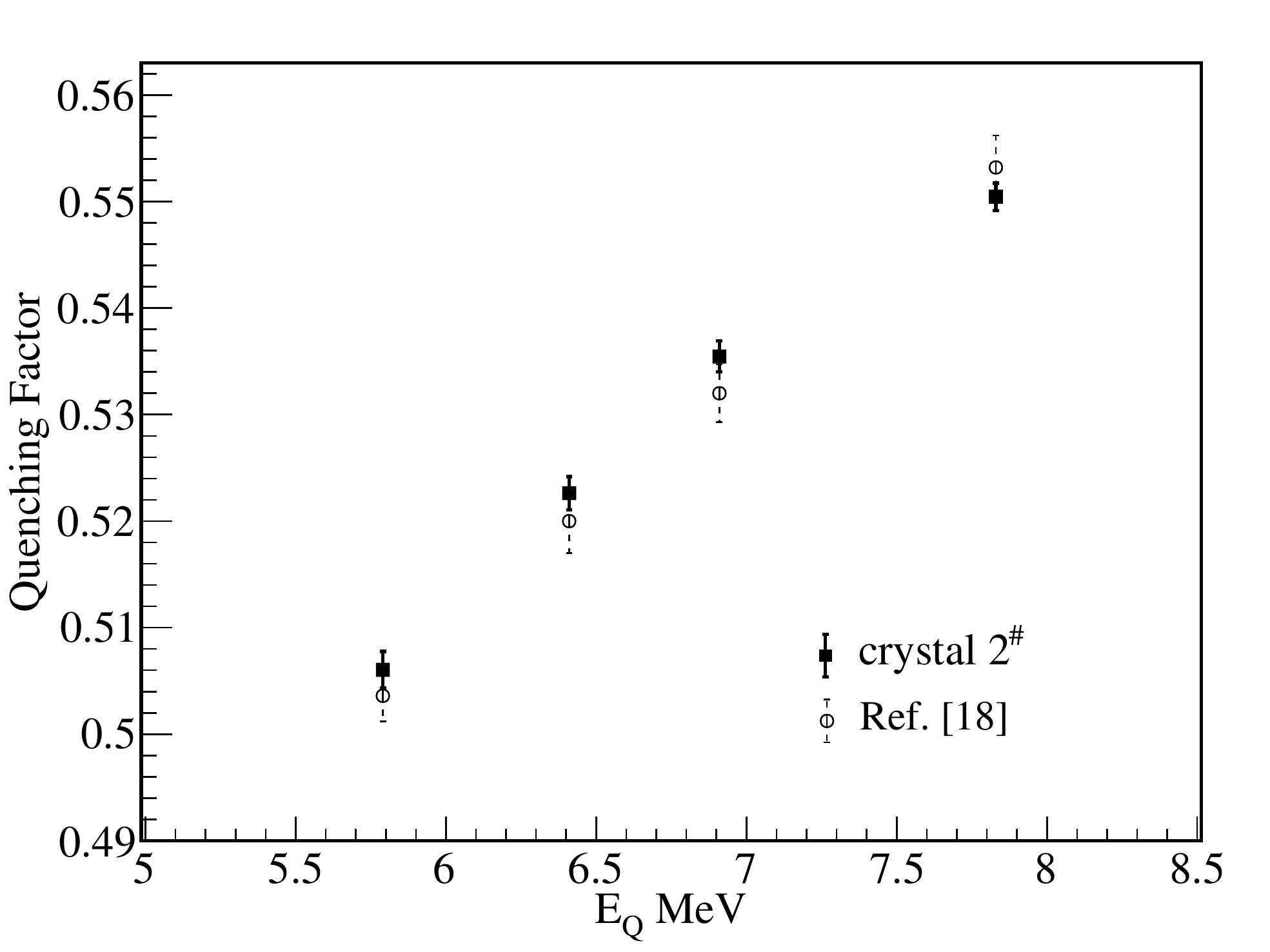}
\figcaption{\label{fig:QF}  The QF result from crystal 2$^{\#}$ consistent with the result from Ref.  \cite{2006_Cs137_contamination}}
\end{center}

The intrinsic contaminations from the $^{137}$Cs, $^{238}$U and $^{232}$Th series are reported, as well as the upper limits of the contamination of $^{235}$U series due to the absence of the correlated events. 
The timing, energy and spatial correlations associated with PSD selection provide significant measurement sensitivities on the contamination levels of U/Th decay chains.

This work provides an event by event selection method to measure the U/Th internal radiopurities and gives complementary references to other experiments.

\acknowledgments{}

\end{multicols}

\centerline{\rule{80mm}{0.1pt}}
\vspace{2mm}

\begin{multicols}{2}

\end{multicols}

\clearpage


\begin{thebibliography}{90}

\vspace{3mm}

\bibitem{2010_Kang} K.J. Kang et al., J. Phys. Conf. Ser, 2010, {\bf 203}: 012028

\bibitem{2012_Yue} Q. Yue and H.T. Wong, J. Phys. Conf. Ser,  2012, {\bf 375}: 042061

\bibitem{2013_1kg} W. Zhao et al., Phys. Rev. D,  2013, {\bf 88}: 052004

\bibitem{2013_Kang} K.J. Kang et al., Chinese Phys. C,  2013, {\bf 37}: 086002

\bibitem{2014_20g} S.K. Liu et al., arXiv: 1403.5421 (2014), Phys. Rev. D, to be submitted for publication

\bibitem{2014_1kg} Q. Yue et al., arXiv: 1404.4946 (2014)

\bibitem{2000_Wong_AP} H.T. Wong et al., Astropart. Phys., 2000, {\bf 14}: 141--152

\bibitem{2012_KIMS_PRL} S.C. Kim et al., Phys. Rev. Letts., 2012, {\bf 108}: 181301

\bibitem{2003_Li_PRL} H.B. Li et al., Phys. Rev. Lett., 2003, {\bf 90}: 131802

\bibitem{2010_Deniz_PRD} M. Deniz et al., Phys. Rev. D, 2010, {\bf 81}: 072001

\bibitem{2013_Kang_FP} K.J. Kang et al., Front. Phys., 2013, {\bf 8}: 412

\bibitem{2013_Wu_CPC} Y.C. Wu et al., Chinese Phys. C, 2013, {\bf 37}: 086001

\bibitem{2004_PSD} S.C. Wu et al., Nucl. Instr. and Meth. A, 2004, {\bf 523}: 116

\bibitem{1964_QF} J.B. Birks., Theory and Practice of Scintillation Counting, Pergamon, London, 1964

\bibitem{1963_QF} R. Gwin, R.B. Murry, Phys. Rev., 1963,  {\bf 523}: 501

\bibitem{2005_Cs137} Y.D. Kim, Nucl. Instr. and Meth. A, 2005,  {\bf 552}: 456--462

\bibitem{2003_Cs137_contamination} T.Y. Kim et al, Nucl. Instr. and Meth. A, 2003,  {\bf 500}: 337--344

\bibitem{2006_Cs137_contamination} Y.F. Zhu et al, Nucl. Instr. and Meth. A, 2006,  {\bf 557}: 490--500

\bibitem{2000_ICPMASS} B.C. Barton et al, Nucl. Instr. and Meth. A, 2000, {\bf 443}: 227

\bibitem{2000_Rn222_PLB} S. Cooper et al, Phys. Lett.B, 2000, {\bf 490}: 6

\bibitem{2010_QF_AP} V.I. Tretyak, Astropart. Phys., 2010, {\bf 33}: 40-53

\end{thebibliography}
\end{document}